\documentclass{epl}
\usepackage{amsmath}

\title{Coulomb-$U$ and magnetic moment collapse in $\delta$-Pu}
\shorttitle{Magnetic moment collapse in $\delta$-Pu}
\author{A. B. Shick\inst{1} \and V. Drchal\inst{1} \and L. Havela\inst{2}}
\institute{
  \inst{1} Institute of Physics, ASCR, Prague, Czech Republic\\
  \inst{2} Charles University, Faculty of Mathematics and Physics, Department of Electronic Structures, Ke Karlovu 5, 121 16 Prague 2, Czech Republic\\
}




\begin{document}

\maketitle

\begin{abstract}
The {\em around-the-mean-field} version of the LDA+U method is applied
to investigate electron correlation effects in $\delta$-Pu.
It yields a non-magnetic ground state of $\delta-$Pu, and provides
a good agreement with experimental equilibrium volume, bulk modulus and
explains important features of the photoelectron spectra.
\end{abstract}

%\section{Introduction}
\vspace*{-0.25cm}
Plutonium is probably the most intricate element
from the point of view of condensed matter physics. It exhibits
six allotropic modifications at ambient pressure, some of them of
very low symmetry (monoclinic), and there is little doubt that the
anomalous behaviour is related to the $5f$ electronic states,
being at the borderline between the localized, non-bonding,
behaviour and the bonding situation of electronic bands. One can
preconceive that at the cross-over regime, several states with
very different degree of $5f$ delocalization can be nearly
degenerated in energy.

The most thoroughly studied phases are $\alpha-$Pu (monoclinic) and
$\delta-$Pu ({\em fcc}).
The latter is stable between 592 and 724 K, but can be stabilized
down to $T = 0$ K by various dopants.
This phase has the largest volume (by 20\% higher than $\alpha-$Pu)
(for overview see Ref. \cite{hecker} and references therein).

The atomic volume is an important indicator of the situation of
the $5f$-electronic states. Withdrawing of the $5f$ states from
the bonding and confining them in the ionic core leads to a
significant volume expansion. On the plot of atomic volumes of
elements, Pu ($\alpha$) represents a continuation of light
actinides, with the decreasing branch resembling the parabolic
behavior of transition metals. On the other hand, heavy actinides,
starting with Am ($Z$ = 95), display higher volume, following a
weakly decreasing volume of lanthanides, characterized by
non-bonding $4f$ states. $\delta-$Pu, being half way between the
volume of $\alpha-$Pu and Am, represents therefore generally the
cross-over regime, where electron-electron correlations play a
prominent role.

{\em Ab-initio} electron energy calculations based on the Density
Functional Theory (DFT) in the Local Density (LDA) or Generalized
Gradient (GGA) approximations account generally well for basic
properties of metallic systems. Numerous variants of this successful
paradigma were applied to Pu phases. The most conspicuous failure is
the case of $\delta-$Pu calculations which inevitably lead to
magnetic ordering. The fact that the lattice expansion due to
magnetism yields approximately correct value of the volume, and
intrasingence of magnetic order within the DFT theory both within
the LDA and GGA approximations, led to speculations about magnetic
ordering of $\delta-$Pu. But it contradicts experimental findings
(magnetic susceptibility has a character of weak Pauli paramagnet
\cite{dormeval03}, paramagnetic state is also evidenced by $^{27}$Al
NMR \cite{fradin}, $^{69}$Ga NMR \cite{Curro04} and neutron
scattering \cite{lashley}). Since its first occurrence
\cite{solov91}, new claims of magnetic Pu keep filling research
journals; there is already a debate about the detailed type of
order, and recently even other Pu phases were assumed as magnetic
\cite{sodsad}. One common feature of light actinides is that orbital
moments, appearing due to the strong spin-orbit interaction in the
case of spin polarization, are oriented antiparallel to spin
moments. LDA or GGA calculations of $\delta-$Pu (with approximately
five $5f$ electrons) lead to the spin moment larger than orbital
moment (which is opposite to what the Hund's rules assume for the
$5f^5$ ionic state), but the existence of "exact" accidental
complete cancellation can be doubted. The systematic cancellation
appears only for the $5f^6$ configuration of Am (both for the $LS$
and $jj$ coupling).

Pu magnetism as an undesirable artefact tends to appear also in
other approaches, like the mixed-scheme model, based on minimization
of energy for the localized $5f^4$ configuration, while the
remaining $5f$ electrons form a band \cite{eriksson99}, while the
Dynamical Mean Field Theory (DMFT) deduces that magnetic order is
washed out by dynamical fluctuations of moments \cite{nature01}. As
yet, the challenging question and the real problem of Pu remains,
which is why is there no magnetism in the $\delta$-Pu phase?

Here, we address this problem making use of a well known L(S)DA+U
method. Generally, the LDA+U calculations account for the on-site
correlations between the $f$ electrons in a more realistic way
than the LSDA. First applied to $\delta-$Pu by Bouchet et al.
\cite{bouchet00}, they also lead to a magnetic solution, similar
to that reported by Savrasov and Kotliar \cite{savrkot00}. In a
contrary to \cite{bouchet00,savrkot00}, we apply to $\delta-$Pu a
different version of the LDA+U method which is based on the original
LDA+U total energy functional of Ref.\cite{AZA91}. We show that
when the LDA+U of Ref.\cite{AZA91} is reformulated in a spin and
orbital rotationally invariant form, it yields a basically
non-magnetic $\delta-$Pu with $S \rightarrow 0$ and $L \rightarrow
0$.

%\section{Methodology}
The correlated band theory L(S)DA+U method
consists of the local spin-density approximation (LSDA) augmented
by a correcting energy of a multiband Hubbard type $E_{ee}$ and a
``double-counting" subtraction term $E_{dc}$ which accounts approximately
for an electron-electron interaction energy already included in
LSDA.
It is well known that the form of ``Coulomb-$U$" correction to
the LDA is not uniquely defined \cite{PMCL03}. The most commonly used
is the version of LDA+U total energy functional in a so-called
``fully localized" limit (FLL) \cite{LAZ95,savrkot00}, in which the
double-counting term $E_{dc}$ is taken to satisfy an atomic-like limit
of the LDA total energy.

Another and historically the first LDA+U functional is often called as an
``around-mean-field" (AMF) limit of the LDA+U.
In this AMF-LDA+U limit \cite{AZA91,CzySaw94} the interaction energy takes the form
\begin{eqnarray}
\label{eq:1}
E_{ee}^{\rm AMF} = \frac{1}{2}  \sum_{\gamma_1, \gamma_2, \gamma_3, \gamma_4}
\delta n_{\gamma_1, \gamma_2}
\Big[ \langle \gamma_1, \gamma_3 |V^{ee}| \gamma_2, \gamma_4 \rangle -
 \langle \gamma_1, \gamma_3 |V^{ee}| \gamma_4, \gamma_2 \rangle \Big]
\delta n_{\gamma_3, \gamma_4} \, ,
\end{eqnarray}
where $V^{ee}$ is an effective on-site ``Coulomb-$U$'' interaction,
the combined spin-orbital index $\gamma = (m \sigma)$ is composed
from the angular $m$ and spin index $\sigma$, and
\begin{equation}
\delta n_{\gamma_1, \gamma_2} = n_{\gamma_1, \gamma_2}
- n^{\sigma_1} \delta_{\gamma_1, \gamma_2} \; , \quad
n^{\sigma} = \frac{1}{2l+1} \sum_{m=-l}^l n_{m \sigma, m \sigma} \; ,
\label{eq:2}
\end{equation}
where $n_{\gamma_1, \gamma_2}$ is the on-site $f$-occupation matrix
in the spin-orbital space which has to be defined with respect to the
chosen localized orbital basis set, and
$n^{\sigma}$
%= \frac{1}{2l+1} \sum_{m=-l}^l n_{m \sigma, m \sigma}$
is an average spin-orbital occupation.
The double-counting correction is set equal to zero, $E_{dc}^{\rm AMF}=0$.

We note that the  essential feature of the
Eq. (\ref{eq:1}) is the presence of spin-off-diagonal
elements of the on-site occupation matrix $n_{\bf \gamma_1
\gamma_2} \equiv n_{m_1 \sigma_1,m_2 \sigma_2}$ which are in
general  non-zero in the presence of the spin-orbit coupling (SOC).

As it was shown recently in Ref.\cite{PMCL03}, the AMF-LDA+U
formulation is most appropriate for metallic phases in which the
Coulomb-$U$ is comparable to the bandwith $W$. On the other hand,
the FLL-LDA+U model is more suitable for insulators, with
Coulomb-$U$  much larger than $W$. Here we have extended the
AMF-LDA+U for the general spin and orbital rotationally invariant
case including the SOC, and implemented it in the full potential
linearized augmented plane wave (FP-LAPW) basis in a way similar
to a previous relativistic FLL-LDA+U FP-LAPW implementation
\cite{SP01}.

%\section{Ground state bulk properties}
We calculated the electronic and magnetic structure of $\delta$-Pu
using three different approaches: (i) LSDA, (ii) FLL-LDA+U, and
(iii) AMF-LDA+U. In all these calculations, we assumed a
tetragonal unit cell with two Pu atoms in order to accommodate the
anti-ferromagnetic (AFM) order, as it is well known that both the
LSDA \cite{KST91} and GGA \cite{soderlind01} yield the AFM as the
ground state of Pu.

\begin{table}[h]
\caption{Ground state spin $M_S$, orbital $M_L$, and total $M_J=M_S+M_L$
magnetic moments (in $\mu_B$) in $\delta$-Pu calculated for
antiferromagnetic configuration at the experimental lattice parameter
$a=$ 8.760 a.u. using the LSDA, FLL-, and AMF-LSDA+U ($U = 4$ eV)
models.
Also given are the equilibrium volume $V_{eq}$ (in (a.u.)$^3$) and bulk
modulus $B$. We note that at the equilibrium the AFM-LSDA is lower in the
total energy than non-magnetic LDA by 12.2 mRy/atom.}
\vspace*{-0.5cm}
\begin{center}
\begin{tabular}{cccccccc}
\hline
\multicolumn{2}{c}{Model}&  $M_S$ & $M_L$ & $M_J$ & $V_{eq}$ & $B$ (kbar) \\
\hline
\multicolumn{2}{c}{LSDA} &4.357 & $-$2.020 & 2.337     & 136.8    & 761    \\
%\multicolumn{2}{c}{LSDA} & 5.057  & 4.357 &$-$2.020 & 2.337  \\
\multicolumn{2}{c}{FLL LSDA+U}& 3.272 &$-$3.802 &$-$0.530 &187.9 & 675     \\
%\multicolumn{2}{c}{FLL LSDA+U} 3.272 &$-$3.802&$-$0.600  \\
\multicolumn{2}{c}{AMF LSDA+U}& $\sim$ 0&$\sim$ 0& $\sim$ 0 & 181.5    & 314     \\
%\multicolumn{2}{c}{AMF LSDA+U}& 5.440  & $\sim$ 0&$\sim$ 0& $\sim$ 0 \\
\hline
\multicolumn{2}{c}{Experiment}& N/A     & N/A   & 0    & 168   & 299 \\
%\multicolumn{2}{c}{Experiment} & 5.4? & N/A     & N/A   & 0    \\
\hline
\end{tabular}
\end{center}
\end{table}

\begin{figure}[t]
\vspace*{-0.5cm}

\twoimages[scale=0.40]{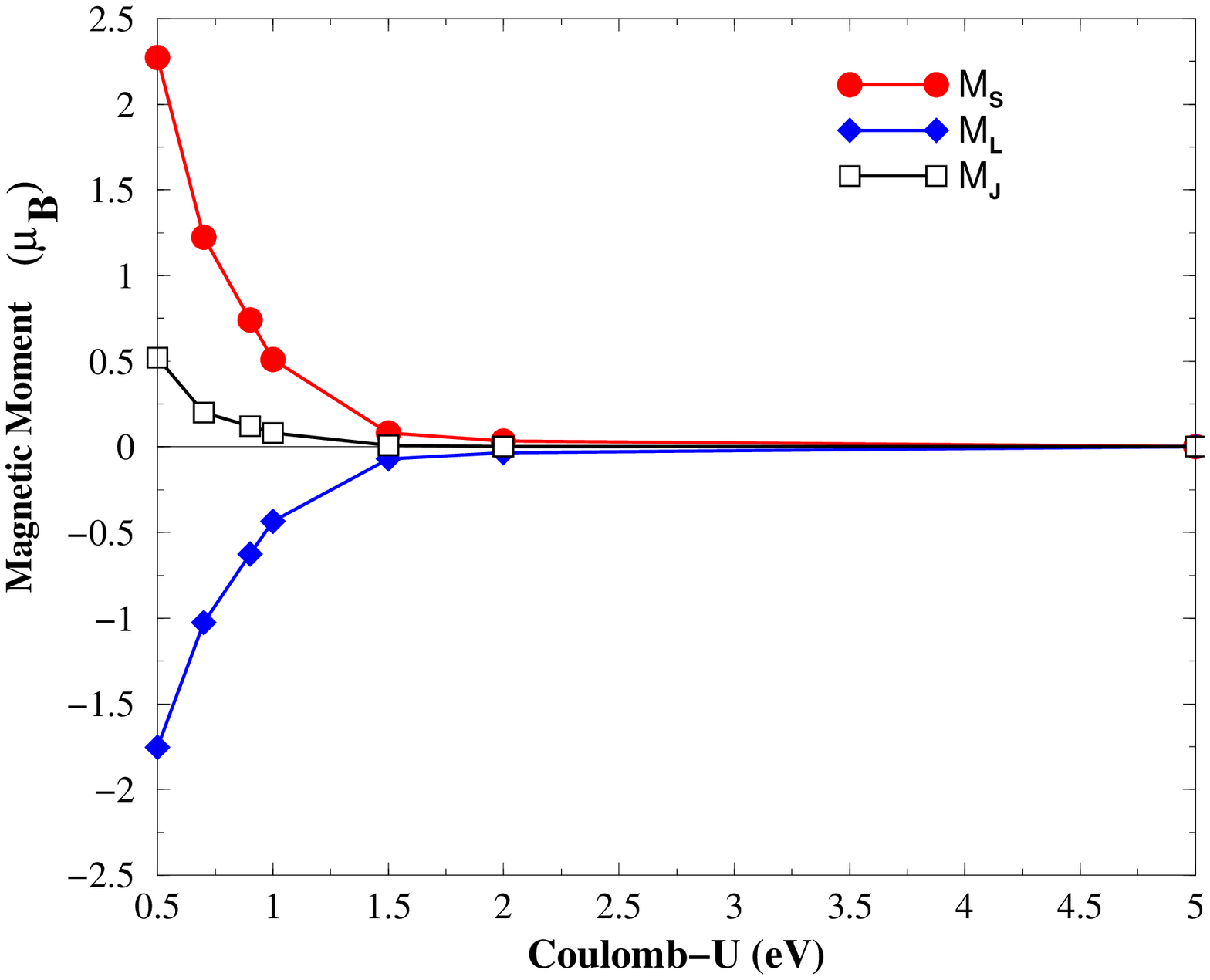}{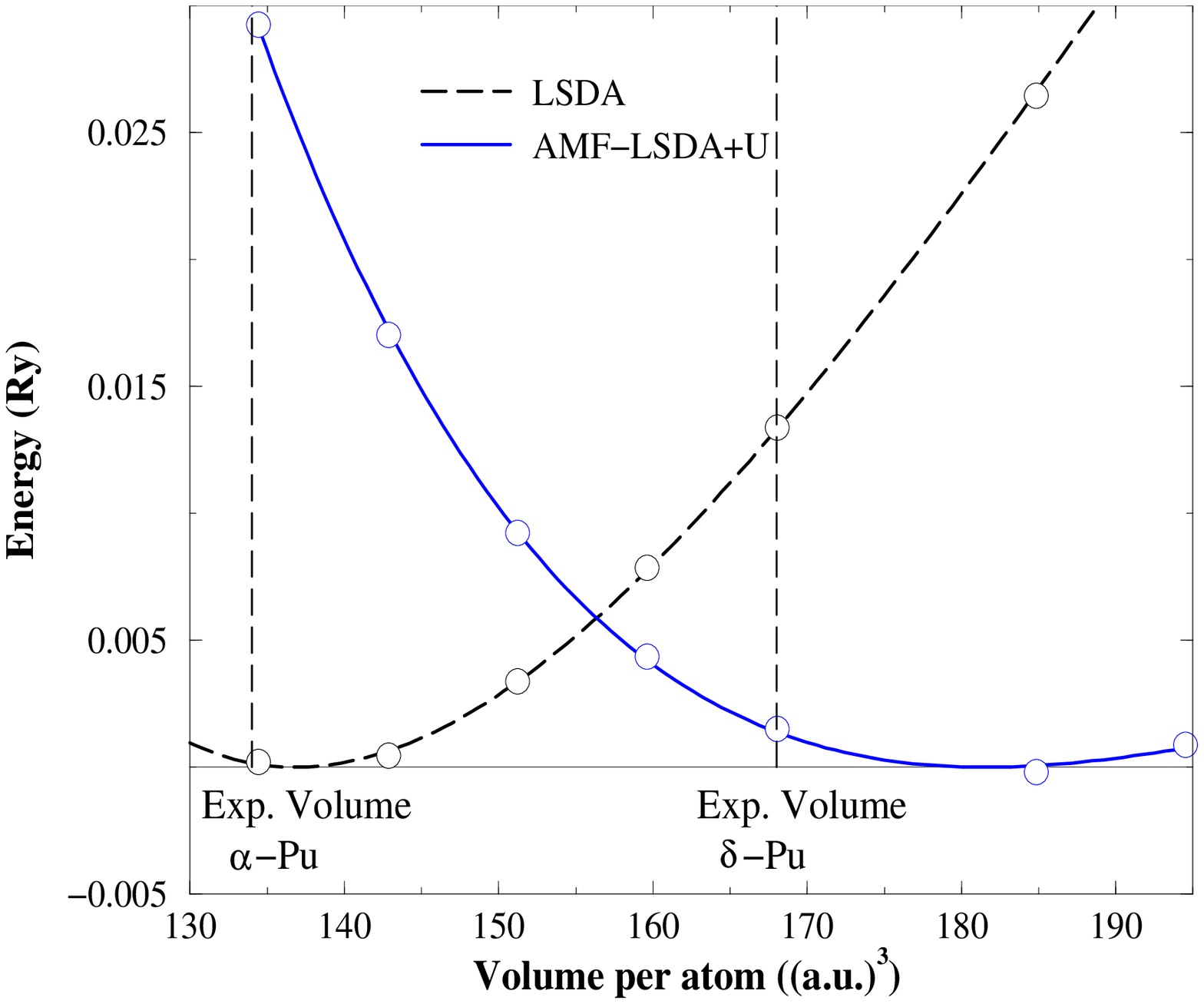}
%\label{fig1}
\caption{Spin $M_S$, orbital $M_L$, and total $M_J$ magnetic moments
dependence on the Coulomb-$U$ as calculated in AMF-LDA+U model for
the experimental lattice parameter $a=$ 8.760 a.u.}
%\label{fig2}
\caption{Total energy as a function of the volume per Pu atom calculated
within the AMF-LSDA+U ($U = 4$, $J=0.7$ eV) and LSDA in the
antiferromagnetic model.}
\end{figure}

The spin $M_S$, orbital $M_L$ and total $M_J$ magnetic moments on
Pu atom calculated within the LSDA (see Table I) are in a very
good agreement with previous LSDA calculations \cite{KST91}.
Starting from the LSDA calculated charge and spin densities and
on-site spin and orbital occupations, we performed the LDA+U
calculations. We choose $U=4$ eV and exchange interaction $J=$ 0.7
eV in the range of commonly accepted values for Pu
\cite{savrkot00}. The FLL-LDA+U model yields a substantial (approx.
1 $\mu_B$) reduction of $M_S$ and an increase of $|M_L|$ by approximately
2 $\mu_B$, resulting in a substantial reduction of the $|M_J|$
(see Table I), while keeping non-zero local magnetic moment on Pu
atom \cite{difference}. The Coulomb-$U$ has a strong effect on
both the spin and orbital polarization, but it changes very little
the total $f$-state occupation $n_f$ -- from approximately 5.06 in
LSDA to 4.95 in FLL-LDA+U. With the increase of $U$ from 4 eV to 5
eV, the $M_S$ increased to 3.42 $\mu_B$ and $M_L$ decreased to
$-3.97 \; \mu_B$, yielding a slight increase of $|M_J|$ to 0.55
$\mu_B$ and a decrease of $n_f$ to 4.93. Thus, in spite of major
differences in the values of the $M_S$, $M_L$, and $M_J$, both the
LSDA and FLL-LDA+U calculations describe $\delta$-Pu as a $5f^5$
state.

Next, we turn to the salient aspect of our investigation, the
AMF-LDA+U ($U=$ 4 eV, $J=$ 0.7 eV) calculations. Starting from
strongly spin-polarized LSDA charge and spin densities and 5$f$
manifold spin and orbital occupations and without any constraint,
the calculations converged to the almost zero magnetic moment with
remaining $M_S$ and $|M_L|$ less than 0.01 $\mu_B$. We also
performed the calculations starting from a different FLL-LDA+U
ground state and obtained essentially the same results. We
conclude that the AMF-LDA+U yields fundamentally non-magnetic
$\delta$-Pu, in accordance with experimental observations.
Importantly, the $5f$ occupation $n_f$ is increased substantially
from $n_f \approx 5$ in LSDA and FLL-LDA+U to $n_f = 5.44$ (see
Table I) meaning that there is a substantial deviation from the
$5f^5$ ionic state.

The dependence of the spin $M_S$, orbital $M_L$, and total $M_J$
calculated within the AMF-LDA+U on the value of Coulomb-$U$ is
shown in Fig. 1. For small values of $U$ ($\approx J$),
$\delta$-Pu is magnetic with sizeable $M_S$ and $M_L$ moments that
almost cancel each other. As the $U$ value is increased to 1.5 eV,
the local moments $M_S$ and $M_L$ disappear and $\delta$-Pu is
non-magnetic for realistic values of the Coulomb-$U$ (from 3 to 5
eV). It is interesting to note that $n_f$ also depends on $U$ and
it increases from $n_f$=5.1 ($U$=0.5 eV) to $n_f$=5.5 ($U$=5 eV).

The total energy dependence on the volume per Pu atom calculated
within the AMF-LDA+U ($U = 4$ eV) is shown in Fig. 2 and compared
with the results of the AFM LSDA calculations. First, we mention
that the AFM LSDA yields the equilibrium volume $V_{eq}$ 136.8
(a.u.)$^3$ per Pu and bulk modulus $B$ = 761 kbar (see Table I),
in a good agreement with the LMTO results of Ref.\cite{KST91}. The
LSDA $V_{eq}$ is close to $\alpha$-Pu value 134 (a.u.)$^3$ which
is approximately by 20 \% smaller than the experimental volume 168
(a.u.)$^3$ of $\delta$-Pu, while the bulk modulus is almost twice
bigger than its experimental value 299 kbar \cite{bmoduli}.
As it was already shown in \cite{bouchet00,savrkot00}, the FLL-LDA+U
model gives the $V_{eq}$ close to the experimental $\delta$-Pu
volume and it slightly lowers the bulk modulus (see Table I).
Finally, the AMF-LDA+U yields $V_{eq}=$ 181.5 eV and $B=$ 314 kbar
(see Table I and Fig. 1) which are in a very good agreement with
experimental data. As yet, the Coulomb-$U$ is treated as an
adjustable parameter and $V_{eq}$ can be fitted further to the
experiment by a slight decrease of $U$. However, it is clear that
the conventional value $U= 4$ eV works already quite well. Thus,
the AMF-LDA+U is shown to give the best agreement with experiment
both for magnetic and structural properties of the $\delta$-Pu.

In Fig. 3(a) we show the total (TDOS, per unit cell) and $f$-states
projected densities of states (fDOS) resulting from the $\delta$-Pu
bulk AMF-LDA+U calculations.
In order to clarify the character of the 5$f$ manifold resulting
from the AMF-LDA+U calculations, it is convenient to look at the
$j_z  = m_l+m_s$ $f$DOS shown in Fig. 3(b), as the $z-$projection
of the total moment $j_z$ is the only remaining quantum number in
the presence of the SOC and magnetic polarization.
It shows six filled $j=5/2$ $f$-states and the $j=7/2$ $f$-states
that are split away by approximately $4-5$ eV.
As the LDA+U model is not based on any kind of atomic coupling
scheme ($LS$ or $jj$) it rather determines the set of single-particle
orbitals that minimize variationally the total energy.
The LDA+U result can be interpreted as yielding the ground-state
configuration that corresponds to the $jj$-coupled Slater determinant
formed of six $j=5/2$ orbitals.
Further, taking into account the finite ($\approx$ 2 eV) width
of the $f$-band and the occupation $n_f$ = 5.44, we can interpret
the calculated ground state as a partially localized $f^6$ manifold
hybridized with a broad valence band (see Fig. 3(a)).

%\begin{figure}
%\oneimage[width=12cm,height=10cm]{fig1.eps}
%\caption{Total energy per atomic volume:  Anti-FM LSDA AMF-LSDA+U= 4 eV}
%\label{fig1}
%\end{figure}

%%%\section{Results and discussion}

%\section{Comparison with PES}

\begin{figure}[t]
\vspace*{-0.5cm}

\oneimage[width=15cm,height=13cm]{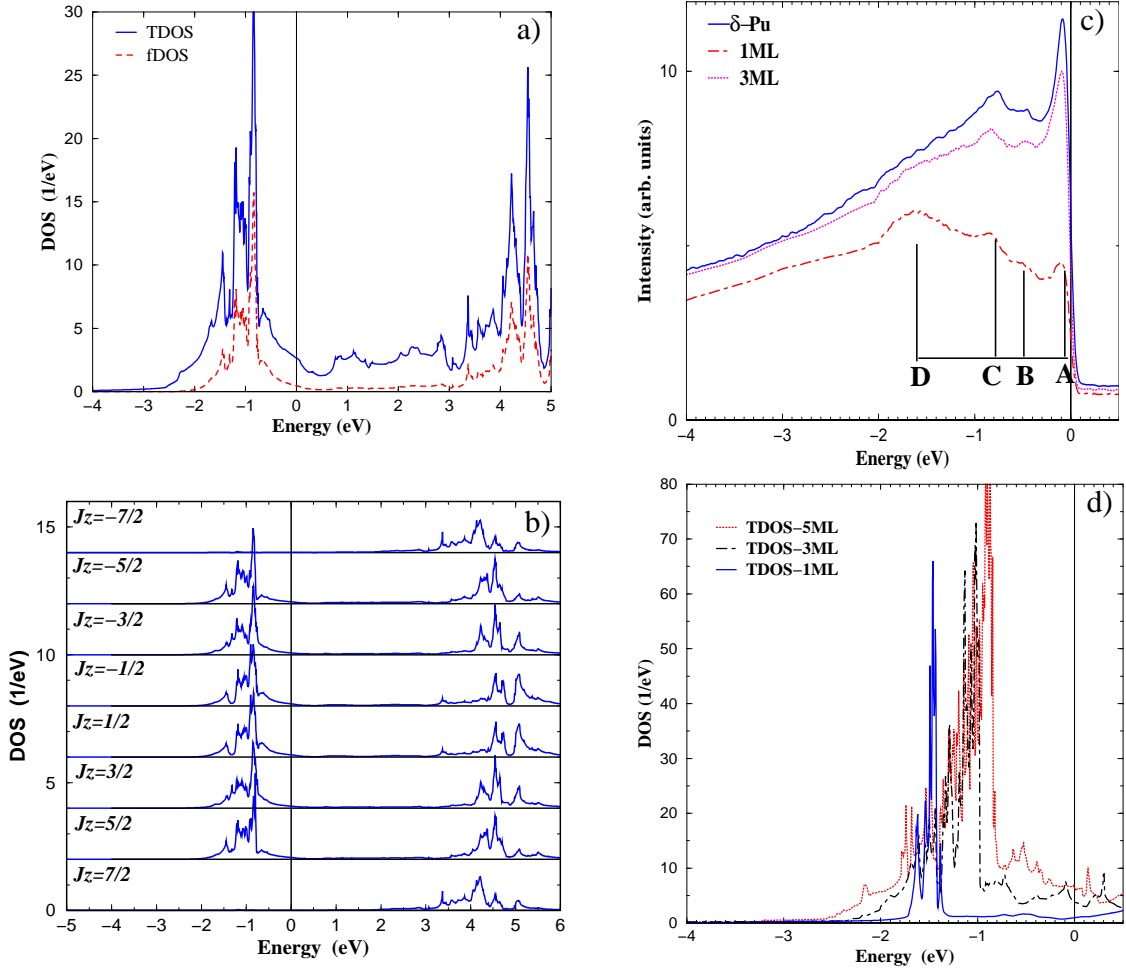}
%\label{fig1}
\caption{a) Total (TDOS) and f-projected (fDOS) DOS as the results
of AMF-LSDA+U=4 eV calculations for the bulk $\delta$-Pu;
b) $\delta-$Pu $j_z-$resolved $f$DOS from
AMF-LDA+U  calculations; c) experimental UPS spectra
for Pu multilayers and the bulk taken
from Ref.\cite{gouder01,havela03}; d) 1ML, 3ML, and 5ML total DOS
calculated for free-standing Pu multilayers making use of
AMF-LDA+U model.}

\end{figure}

An important criterion for assessment of results of electronic
structure calculations is given by electron spectroscopies.
Although some techniques, as BIS, have not been applied on Pu yet,
there exists high-resolution Ultraviolet Photoelectron
Spectroscopy (UPS) data, mapping the electronic structure down to
about 10 eV below $E_{\rm F}$
\cite{arko00,gouder01,havela02,terry02},
%%(Arko et al. PRB 2000,
%%Gouder et al. Europhys.Lett. 2001, Havela et al. PRB 2002,
%%Terry et. al., Surf.Sci. 2002)
while X-ray photoelectron spectroscopy
\cite{gouder01,havela02,terry02}
%%(XPS Gouder et al. Europhys.Lett. 2001, Havela et al. PRB 2002,
%%Terry et. al., Surf.Sci. 2002)
gives information on the screening of a deep core-hole by conduction
electrons.

Valence-band spectra of $\delta-$Pu and handful of other Pu-based
systems studied so far exhibit invariably three narrow features,
one at $E_{\rm F}$ or in its close vicinity, the other two at 0.5
and 0.85 eV below $E_{\rm F}$, which are called A, B, and C,
respectively. Some systems exhibit also another, broader,
separate feature D, the energy of which varies between compounds,
and which is not noticeable for $\delta-$Pu. It was argued
\cite{havela02} that the intensity of D reflects the degree of
5$f$ localization. For example, the peaks A-C are absent in the
5$f$ localized compound PuSb \cite{gouder00}, and all the 5$f$
spectral intensity is concentrated in the broad peak D.

Valence-band photoelectron spectra reflect to some extent the density
of occupied states in the ground state (DOS).
A comparison with the bulk experimental UPS \cite{gouder01,havela02}
(see Fig. 3(c)) shows that LDA+U places the $5f$ manifold
approximately 0.9 eV below the Fermi energy ($E_{\rm F}$) in accordance
with the experimental C peak position, while it does not resolve correctly
the experimental features A and B at the $E_{\rm F}$ edge.

As discussed in Ref. \cite{havela03}, the sharp features A-C are
not specific for individual compounds, but they appear at
invariable energies over a class of diverse Pu systems.
It is therefore implausible that they could be related to any
ground-state one electron states. Dismissing the peaks A-C from the
comparison, as we assume them either to be due to the final-state
effects, which could be reconciled with the $5f^5$ final-state
multiplet \cite{gerken83}, or due to a general many body
phenomenon (analogy of a Kondo effect) \cite{gunnsch83}), we are
left at a difficult situation, as no other well-defined feature
is seen in the UPS spectra. But for ultrathin layers, with
thickness varying down to about one monolayer (see Fig. 3(c)), a
broad, but well defined peak D, centered at 1.6 eV binding energy,
concentrates most of the $5f$ spectral intensity, whereas the
peaks A-C (still at their energies) become barely noticeable.

Therefore we undertook analogous AMF-LDA+U calculations for
free-standing Pu slabs (where the number of Pu-layers varies from 1
to 5) with  the lattice parameter identical to that of $\delta-$Pu.
For the one monolayer (1ML) (see Fig. 3(d)), the $5f$ states form a
very narrow band at 1.5 eV, which is in an excellent agreement with
the position of the D-peak (cf. Fig 3(c)). An intermediate
situation, Pu trilayer (3ML), demonstrates that increasing thickness
leads to the band broadening, but also to a shift towards $E_{\rm
F}$. With a further increase of the Pu slab thickness up to five
monolayers (5ML), the band broadening continues and the DOS becomes
very similar to the bulk (cf. Fig. 3(a)). This broader $5f$ band is
in fact underlying the peak C, which explains why it is not
observed. Moreover, the fact that in all four cases (bulk, 1, 3, and
5 monolayers) the non-magnetic state $(L = 0\, ,\, S = 0)$ persists,
confirms the robustness of this solution.

It is also to mention that both LSDA and FLL-LDA+U magnetic solutions
yield the DOS which does not correspond to the experimental UPS spectra
shown in Fig. 3(c).
In the LSDA case, the 5$f$-band has a very sharp peak structure in the
region from $-1.2$ eV to 0.5 eV around $E_{\rm F}$.
For the FLL-LDA+U, the 5$f$-band becomes localized with the 5$f$ manifold
well separated below the bottom of the valence band in the energy region
from 2.5 to 4.5 eV below the $E_{\rm F}$, again in a contradiction to
experimental data.

%\section{Character of 5$f$ states in the $\delta$-Pu}
%\begin{figure}[t]
%\vspace*{-1cm} \oneimage[width=8cm,height=7cm]{3jzAMF.eps}
%\label{fig4} \caption{The $\delta-$Pu $j_z-$resolved $f$-DOS from
%AMF-LDA+U ($U=4$ eV) calculations. }
%\end{figure}

To summarize, we have shown that the {\em around-the-mean-field}
version of the LDA+U method yields non-magnetic ground state of
$\delta-$Pu and, at the same time, it gives a good agreement with
experimental data, namely, with the photoemission spectra,
equilibrium volume, and bulk modulus.
%ned addition
The fact that the non-magnetic character is not due to a mutual
cancellation of spin and orbital parts of the moment, but is due to
$S = 0$ and $L = 0 $ , is particularly corroborated by very recent
neutron scattering experiments \cite{lashley}. We are aware that our
approach does not account for dynamical effects, which may become
very important, as suggested in Ref.\cite{nature01}. Also, as we
focused here on the magnetic, structural and  spectroscopic
properties of the $\delta-$Pu only, we do not discuss the $\alpha-$
to $\delta-$Pu phase transition, leaving it for further
consideration.

%\acknowledgments
The research was carried out by A.B.S. and V.D. within the project
AVOZ1-010-914 of the Academy of Sciences of the Czech Republic.
Financial support was provided by the Grant Agency of the Academy of
Sciences of the Czech Republic (Project A1010203) and by the Czech
Science Foundation (Project 202/04/1103). This work was also a part of the
research program MSM0021620834 financed by the Ministry of Education
of the Czech Republic. We gratefully acknowledge valuable discussion
with A. I. Lichtenstein.


\vspace*{-0.25cm}
\begin{thebibliography}{0}

\bibitem{hecker}
  \Name{Hecker S.S., Harbur D.R. \and Zocco T.G.}
  \REVIEW{Progr. Materials Sci.}{49}{2004}{429}.

\bibitem{dormeval03}
  \Name{Dormeval M., Baclet N., Valot C., Rofidal P. and Fournier J.M.}
  \REVIEW{J. Alloys Comp.}{350}{2003}{86}.

\bibitem{fradin}
  \Name{Fradin F.Y. \and Brodsky M.B.}
  \REVIEW{Int. J. Magnetism}{1}{1970}{1}.

\bibitem{Curro04}
  \Name{Curro N.J. \and Morales L.}
  \REVIEW{arXiv:cond-mat/0404626}{}{2004}.

\bibitem{lashley}
  \Name{Lashley J.C., Lawson A., McQueeney R.J., \and Lander G.H.}
% \REVIEW{Phys. Rev. B}{}{to be published}{}.
  \REVIEW{arXiv:cond-mat/0410634}{}{2004}.

\bibitem{solov91}
  \Name{Solovyev I.V., Liechtenstein A.I., Gubanov V.A., Antropov V.P.
        \and Andersen O.K.}
  \REVIEW{Phys. Rev. B}{43}{1991}{14414}.

\bibitem{sodsad}
  \Name{S\"oderlind P. \and Sadigh B.}
  \REVIEW{Phys. Rev. Lett.}{92}{2004}{185702}.

\bibitem{eriksson99}
  \Name{Eriksson O., Becker J.D., Balatsky A.V. \and Wills J.M.}
  \REVIEW{J. Alloys Comp.}{287}{1999}{1}.

\bibitem{nature01}
  \Name{Savrasov S.Y., Kotliar G. \and Abrahams E.}
  \REVIEW{Nature}{410}{2001}{793}.

\bibitem{bouchet00}
  \Name{Bouchet J., Siberchicot B., Jollet F. \and Pasturel A.}
  \REVIEW{J. Phys.: Condens. Mater.}{12}{2000}{1723}.

\bibitem{savrkot00}
  \Name{Savrasov S.Y. \and Kotliar G.}
  \REVIEW{Phys. Rev. Lett.}{84}{2000}{3670}.

\bibitem{AZA91}
  \Name{Anisimov V.I., Zaanen J.,
        \and Andersen O.K.}
  \REVIEW{Phys. Rev. B}{44}{1991}{943}.
  {The LDA+U method can be regarded as the Hartree-Fock limit of the
   LSDA combined with the dynamical mean field theory (LSDA+DMFT) \cite{nature01}.}

\bibitem{PMCL03}
  \Name{Petukhov A. G., Mazin I.I., Chioncel L., Liechtenstein A.I.}
  \REVIEW{Phys. Rev. B}{67}{2003}{153106}.

\bibitem{LAZ95}
  \Name{Liechtenstein A.I., Anisimov V.I. \and Zaanen J.}
  \REVIEW{Phys. Rev. B}{52}{1995}{R5467}.

\bibitem{CzySaw94}
  \Name{Czy\.zyk M.T. \and Sawatzky G.A.}
  \REVIEW{Phys. Rev. B}{49}{1994}{14211}.

\bibitem{SP01}
  \Name{Shick A.B. and Pickett W.E,}
  \REVIEW{Phys. Rev. Lett.}{86}{2001}{300};
  \Name{Shick A.B., Liechtenstein A.I., and Pickett W.E}
  \REVIEW{Phys. Rev. B}{60}{1999}{10763}.

\bibitem{KST91}
\Name{Katsnelson M.I., Solovyev I.V., and Trefilov A.V.}
\REVIEW{JETP Letters}{56}{1992}{272}.

\bibitem{soderlind01}
  \Name{S\"oderlind P.}
  \REVIEW{Europhys. Lett.}{55}{2001}{525}.

\bibitem{difference}
We note that our FLL-LDA+U results with assumed AFM order differ somewhat
from those previously reported  in Refs. \cite{bouchet00,savrkot00}
which were obtained in the basis of linear-muffin-tin orbitals (FP-LMTO)
assuming a ferromagnetic order.
It is also to say that the results of \cite{bouchet00} and \cite{savrkot00}
differ from each other, in spite of the use of the same FP-LMTO package
of S. Savrasov \cite{savrkot00}.

\bibitem{bmoduli}
\Name{Ledbetter H.M. and Moment R.L.}
  \REVIEW{Acta Metall.}{24}{1976}{891}.

\bibitem{gouder01}
  \Name{Gouder T., Havela L., Wastin F. \and Rebizant J.}
  \REVIEW{Europhys. Lett.}{55}{2001}{705}.

\bibitem{havela03}
  \Name{Havela L., Wastin F., Rebizant J. \and Gouder T.}
  \REVIEW{Phys. Rev. B}{68}{2003}{085101}.

\bibitem{arko00}
  \Name{Arko A.J., Joyce J.J., Morales L., Wills J., Lashley J.,
        Wastin F. \and Rebizant J.}
  \REVIEW{Phys. Rev. B}{62}{2000}{1773}.

\bibitem{havela02}
  \Name{Havela L., Gouder T., Wastin F. \and Rebizant J.}
  \REVIEW{Phys. Rev. B}{65}{2002}{235118}.

\bibitem{terry02}
  \Name{Terry J. {\em et al.,}}
% Schulze R.K., Farr J.D., Zocco T., Heinzelman K.,
%        Rotenberg E., Shuh D.K., Van der Laan G., Arena D.A. \and
%        Tobin J.G.}
  \REVIEW{Surf. Sci.}{499}{2002}{L141}.

\bibitem{gouder00}
  \Name{Gouder T., Wastin F., Rebizant J. \and Havela L.}
  \REVIEW{Phys. Rev. Lett.}{84}{2000}{3378}.

\bibitem{gerken83}
  \Name{Gerken F. \and Schmidt-May J.}
  \REVIEW{J. Phys. F: Met. Phys.}{13}{1983}{1571}.

\bibitem{gunnsch83}
  \Name{Gunnarson O., and Sch\"onhammer K.}
  \REVIEW{Phys. Rev. B}{28}{1983}{4315}.


%\bibitem{remark2}
%One has to keep in mind that $n_f$ is not uniquely
%defined and depends on the choice of the ``muffin-tin" (MT) radius.
%It is to note that with the reasonable MT-radius choise from 2.85 to 3.0 a.u.,
%this dependence is rather weak due to 5$f$ states spatial localization.

\end{thebibliography}
\end{document}